\documentclass[conference, a4paper]{IEEEtran}
	\newtheorem{theorem}{Theorem}
	\newtheorem{lemma}[theorem]{Lemma}
	\newtheorem{definition}[theorem]{Definition}

\usepackage[cmex10]{amsmath}
\usepackage{mathtools, amssymb}
	\def\degrade{\mathrel{-\mkern-2mu\triangleright}}
	\def\>{\succcurlyeq}
	\def\<{\preccurlyeq}
	\def\properscale#1#2{\raisebox{2.9pt}%
	{\kern1pt\scalebox{#1}{\kern-1pt\raisebox{-2.5pt}{#2}}}}
	\def\tbm>{\mathrel{\ooalign{%
		$\succcurlyeq$\cr
		\properscale{.8}{$\succ$}\cr
		\properscale{.64}{$\succ$}\cr
		\properscale{.512}{$\succ$}\cr
		\properscale{.4096}{$\succ$}\cr
		\properscale{.32768}{$\succ$}\cr
		\properscale{.262144}{$\succ$}\cr
		\properscale{.2097152}{$\succ$}\cr
	}}}
	\def\ZZ{{\mathbf Z}}
	\def\QQ{{\mathbf Q}}
	\def\CC{{\mathbf C}}
	
	\def\KK{{\mathbf K}}
	\def\AAFF{{\mathbf{AF}}}
	\DeclareMathOperator\Frac{Frac}
	\DeclareMathOperator\Lindenmayer{Lindenmayer}

\usepackage{xcolor}

\usepackage[noadjust]{cite}
\usepackage[unicode]{hyperref}
	\hypersetup{
		colorlinks, allcolors=green!20!blue!80!black,
		pdfsubject={Information Theory (cs.IT)},
		pdfkeywords={
			polar code,
			partial order,
			inverse density evolution,
			Galois theory,
			real-root isolation,
		},
	}

\begin{document}

\title{Density Devolution for Ordering Synthetic Channels}

\author{%
	\IEEEauthorblockN{Hsin-Po Wang}%
	\IEEEauthorblockA{%
		University of California, Berkeley, CA, USA\\
		simple@berkeley.edu%
	}
	\and
	\IEEEauthorblockN{Chi-Wei Chin}%
	\IEEEauthorblockA{%
		Apricob Biomedicals Co Ltd, Taiwan\\
		qin@quantumsafe.dev%
	}
}
\maketitle

\begin{abstract}\boldmath\def\>{\succcurlyeq}%
	Constructing a polar code is all about selecting a subset of rows from a
	Kronecker power of $[^1_1{}^0_1]$.  It is known that, under successive
	cancellation decoder, some rows are Pareto-better than the other.  For
	instance, whenever a user sees a substring $01$ in the binary expansion of a
	row index and replaces it with $10$, the user obtains a row index that is
	always more welcomed.  We call this a ``rule'' and denote it by $10 \> 01$.
	In present work, we first enumerate some rules over binary erasure channels
	such as $1001 \> 0110$ and $10001 \> 01010$ and $10101 \> 01110$.  We then
	summarize them using a ``rule of rules'': if $10a \> 01b$ is a rule, where
	$a$ and $b$ are arbitrary binary strings, then $100a \> 010b$ and $101a \>
	011b$ are rules.  This work's main contribution is using field theory,
	Galois theory, and numerical analysis to develop an algorithm that decides
	if a rule of rules is mathematically sound.  We apply the algorithm to
	enumerate some rules of rules. Each rule of rule is capable of generating an
	infinite family of rules.  For instance, $10c01 \> 01c10$ for arbitrary
	binary string $c$ can be generated.  We found an application of $10c01 \>
	01c10$ that is related to integer partition and the dominance order therein.
\end{abstract}

\section{Introduction}

	Ever since polar code was invented \cite{Ari09}, it has constantly reminded
	researchers about its similarity to Reed--Muller codes \cite{Ari08, Ari10,
	MHU14, LST14, AbY20} for that they both are generated by some subsets of
	rows of a Kronecker power of $[^1_1{}^0_1]$.

	There is, nevertheless, a huge discrepancy.  For Reed--Muller code, a user
	always picks rows with the highest Hamming weights up to the desired rate.
	Whether this will yield a good code is a very challenging problem
	\cite{KKM17, ReP21}.  For polar code, the story goes the opposite way.
	While Arıkan proved the existence of subsets of rows that are
	capacity-achieving under successive cancellation decoder, it was not clear
	if there is any simple rule that tells us what to choose other than actually
	simulating the channels and carrying out the statistics.  Should we be not
	careful and choose the wrong row, the performance will go low.
	
	Mori and Tanaka \cite{MoT09}, to address the row-selection issue, suggested
	using density evolution and the degradation relation $W^1 \degrade W
	\degrade W^0$ as an alternative to simulation.  Degradation does not
	directly tell us which rows are to select, but it tells us that some rows
	are always better than the other.  This means that not all rows need to
	undergo simulation or density evolution, presumably simplifying the
	row-selection process.

	Later, Schürch \cite{Sch16} and Bardet, Drägoi, Otmani, and Tillich
	\cite{BDO16} added one more degradation relation $W^{10} \degrade W^{01}$.
	Mondelli and Hassani and Urbanke \cite{MHU19} then showed that the number of
	rows that need to be seriously evaluated (either by simulation or evolution)
	is sub-linear in the block length.  For the majority of rows, the
	degradation relation alone is enough to make the decision.
	
	In this work, we follow the track of Wu and Siegel \cite{WuS19}, Camps,
	López, Matthews, and  Sarmiento \cite{CLM21}, Drägoi and Cristescu
	\cite{DrC21}, Geiger \cite{Gei18}, Kahraman \cite{Kah17}, and Kim, Oh, Kim,
	and Ha \cite{KOK16} to study comparison between rows.  Throughout the work,
	we assume that the underlying channel is a binary erasure channel (BEC)
	and the successive cancellation decoder is in use.
	
	For the first time, we consider \emph{density devolution}, which stands for
	the opposite operation of density evolution.  In the absence of devolution,
	comparing two BECs is essentially about proving a certain polynomial
	nonnegative, which can be done procedurally.  Devolution introduces
	square root into the world of polynomials and the problem become proving
	certain algebraic function nonnegative.  Wielding tools from field theory,
	Galois theory, and numerical analysis, we can decide, rigorously, if an
	algebraic function is always nonnegative.  Each nonnegative algebraic
	function imply the nonnegativity of infinitely many polynomials, which means
	that we can now prove polynomial inequalities en masse.

	The paper is organized as follows.
	Section~\ref{sec:code} reviews the definition of
		polar code and a related preorder.
	Section~\ref{sec:2ary} shows how to determine a polynomial inequality.
	Section~\ref{sec:devolve} extends the alphabet to $\{0, 1, 2, 3\}$.
	Section~\ref{sec:field} lays some field theory bricks.
	Section~\ref{sec:4ary} shows how to determine
		a inequality involving $\{0, 1, 2, 3\}$.
	Section~\ref{sec:tbm} proves infinitely many inequalities.

\section{Polar Code and Partial Order over BEC}\label{sec:code}

	A simplified recipe of polar code reads: Let $K \coloneqq [^1_1{}^0_1]$.
	Choose a positive integer $n$ and construct $K^{\otimes n}$, the $n$th
	Kronecker power of $K$.  Then, choose a subset of rows of $K^{\otimes n}$ as
	the generator matrix $G$.  This $G$ generates a polar code.

	In selecting rows, a useful fact is that each row corresponds to a synthetic
	channel that is defined through density evolution.  Let's take the BEC with
	capacity $x \in [0, 1]$ as an example.  First, define $I_0(x) \coloneqq x^2$
	and $I_1(x) \coloneqq 1 - (1 - x)^2$.  Next, define $I_{b_1 b_2 \dotsm b_n}
	(x) \coloneqq I_{b_2 \dotsm b_n} (I_{b_1} (x))$ for any binary string $b = b_1
	b_2 \dotsm b_n \in \{0, 1\}^n$.  Now, the $(1 + \sum_{i=1}^n b_i 2^{n-i})$th
	row of $K^{\otimes n}$ corresponds to the BEC with capacity $I_b(x)$.

	The design principle of polar code recommends that the best rows are those
	with the highest capacities.  A natural question is: Can there be two rows
	such that one always has a higher capacity than the other?

	\begin{definition}
		For any binary strings $a$ and $b$, we say $a \> b$ if $I_a(x) \geq
		I_b(x)$ for all $x \in [0, 1]$.
	\end{definition}

	The following lemmas are known regarding $I$ and $\>$.

	\begin{lemma}
		$I_0$ and $I_1$ are strictly monotonically increasing and map $[0, 1]$
		bijectively onto $[0, 1]$.  In fact, all $I_b$ are.
	\end{lemma}

	\begin{IEEEproof}
		Being strictly monotonically increasing and bijective is closed under
		function composition.
	\end{IEEEproof}

	\begin{lemma}\label{lem:trans}
		$\>$ forms a preorder on binary strings.
	\end{lemma}

	\begin{IEEEproof}
		Reflexivity: $I_a(x) \geq I_a(x)$, hence $a \> a$.  Transitivity:
		$I_a(x) \geq I_b(x) \geq I_c(x)$ implies $I_a(x) \geq I_c(x)$.
	\end{IEEEproof}

	\begin{lemma}\label{lem:mono}
		$ac \> bd$ if $a \> b$ and $c \> d$ for any binary strings $a$, $b$,
		$c$, and $d$.  Juxtaposition stands for string concatenation.
	\end{lemma}

	\begin{IEEEproof}
		$I_c$ is monotonically increasing so $ac \> bc$.  The image $I_b([0,
		1])$ is contained in $[0, 1]$ so $bc \> bd$.
	\end{IEEEproof}

	\begin{lemma}\label{lem:dual}
		If $a \> b$, then $\tilde b \> \tilde a$.  Tilde stands for bitwise
		complement.
	\end{lemma}

	\begin{IEEEproof}
		The plot of $I_{\tilde a}$ is the plot of $I_a$ rotated by $180$ degrees
		w.r.t.\ $(1/2, 1/2)$ as the center.  Hence the assumption that $I_a$'s
		plot lies above $I_b$'s plot implies that $I_{\tilde a}$'s plot lies
		below $I_{\tilde b}$'s plot.
	\end{IEEEproof}

	These lemmas together with $1 \> 0$ imply comparisons such as $1111 \> 1011
	\> 1010 \> 0010 \> 0000$.  Readers might argue that this should not be
	counted as multiple contributions.  Rather, it is but one simple rule, $1 \>
	0$, applied multiple times.  Likewise, $1100 \> 1010 \> 1001 \> 0101 \>
	0011$ can be summarized by a simple rule, $10 \> 01$.

	The main goal of this paper is to demonstrate how to generate new rules that
	are not consequences of other rules.  We hope readers understand that what
	qualifies as an independent rule is context-dependent, sometimes even
	subjective.  For instance, $0011 \> 1000$ probably looks like an independent
	rule, but it is a consequence\footnote{ Take the bitwise complement: $011 \>
	10$ implies $01 \> 100$.  Hence $0011 \> 010 \> 1000$.} of $011 \> 10$.  For
	longer comparisons such as $01011 \> 10100$, it is never clear if they can
	be verified without brute force.  Therefore, we will declare a new rule
	whenever it seems to be so.
	
\section{Polynomial Inequality Problem}\label{sec:2ary}

\subsection{A procedure}

	Suppose $P \in \ZZ[x]$ is an polynomial with integer coefficients.  There
	exist algorithms that find the \emph{exact} number of roots of $P$ in a
	given interval (cf.\ Sturm's theorem) or at least a mathematical sound upper
	bound on the number of roots (cf.\ Vincent's theorem) \cite{SaM16}.  A
	root-counting algorithm is usually paired with a bisecting strategy and the
	Newton--Raphson method to pinpoint the roots.  But for now, we only need the
	counting part.

	Assume $P \in \ZZ[x]$.  Here is how to decide whether
	\begin{equation}
		P(x) \geq 0 \qquad \text{for all }x \in [0, 1].
		\label{ine:polys}
	\end{equation}
	Step one: Divide $P$ by $x^m (1-x)^n$, where $m$ and $n$ are the
	multiplicities of $0$ and $1$, respectively, as roots of $P$.  Since $x$ and
	$1 - x$ are nonnegative over $[0, 1]$, the sign of $P$ stays unchanged after
	division.
	
	Step two: Remove the perfect square part of $P$ by dividing $P$ by $(g /
	\gcd(g, g'))^2$, where $g \coloneqq \gcd(P, P')$ (cf.\ Yun's algorithm
	\cite{Yun76}).  To see why this division works, suppose $d \in \ZZ[x]$ is
	any irreducible factor of $P$.  Write $d^m \mathrel\Vert P$ to mean $d^m
	\mid P$ but $d^{m+1} \nmid P$, i.e, $m$ is the multiplicity of $d$ in the
	factorization of $P$.  We then have
	\begin{align*}
		d^{\max(0, m-1)} & \mathrel\Vert P', \\
		d^{\max(0, m-1)} & \mathrel\Vert g, \\
		d^{\max(0, m-2)} & \mathrel\Vert \gcd(g, g'), \\
		d^{\max(0, m-1) - \max(0, m-2)} & \mathrel\Vert g / \gcd(g, g').
	\end{align*}
	Note that $\max(0, m-1) - \max(0, m-2)$ is just the indicator of $m \geq 2$.
	This means that we are removing two copies of $d$ from $P$ if there are two
	or more copies.  Repeat this step several times until the multiplicities of
	all irreducible factors become $0$ or $1$.  Since we always divide $P$ by a
	perfect square, the sign of $P$ stays unchanged.

	Step three: Granted that $P$ is now square-free, we evaluate $P(1/2)$.  If
	$P(1/2) < 0$, this violates \eqref{ine:polys} right away.  If $P(1/2) = 0$,
	since $x = 1/2$ is a single root, either $P(1/2 + \varepsilon)$ or $P(1/2 -
	\varepsilon)$ will be negative for some small $\varepsilon > 0$, violating
	\eqref{ine:polys}.  In either case, our procedure terminates early and
	returns FALSE.

	Step four: Count the number of roots of $P$ in the interval $[0, 1]$.  If
	there is any root, say $x = \xi$, then by the same reasoning as above,
	$P(\xi + \varepsilon)$ or $P(\xi - \varepsilon)$ will be negative, violating
	\eqref{ine:polys}.  If there are no roots, then all $P(x)$ share the same
	sign as $P(1/2)$, confirming that \eqref{ine:polys} is TRUE.  This concludes
	our procedure to decide \eqref{ine:polys}.

	\begin{theorem}
		The procedure described in this subsection decides inequalities of the
		form \eqref{ine:polys}.
	\end{theorem}

\subsection{Application of the procedure}

	Checking $I_a(x) \geq I_b(x)$ is equivalent to checking $I_a(x) - I_b(x)
	\geq 0$.  This constitutes a terminating procedure that decides if $a \> b$
	for any binary strings $a$ and $b$.  One runs this procedure and finds the
	following comparisons.
	\begin{itemize}\small
		\item $1 \> \epsilon \> 0$, where $\epsilon$ is the empty string.
		\item $11 \> 10 \> 01 \> 00$.
		\item $011 \> 10 \> 01 \> 100$.
		\item $111 \> 110 \> 101 \> 011 \> 100 \> 010 \> 001$.
		\item $1111 \> 1110 \> 1101 \> 1011 \> 0111 \> 1100 \> 1010 \> 1001
		    \> 0110 \> 0101 \> 0011 \> 1000 \> 0100 \> 0010 \> 0001 \> 0000$.
		\item $11111 \> 11110 \> 11101 \> 11011 \> 10111 \> 01111 \> 11010$.
		\item $10111 \> 11100 \> 11010 \> 11001 \> 10110
			\> 10101 \> 10011 \> 01101 \> 11000 \> 10100$.
		\item $10101 \> 01110 \> 01101 \> 01011
			\> 10100 \> 10010 \> 10001 \> 01010$.
		\item $01011 \> 00111 \> 10010 \> 01100 \> 01010
			\> 01001 \> 00110 \> 00101 \> 00011 \> 01000$.
		\item $00101 \> 10000 \> 01000 \> 00100 \> 00010 \> 00001 \> 00000$.
	\end{itemize}

	Among these comparisons, $1 \> \epsilon$\,,\, $011 \> 10 \> 01$\,,\, $1001
	\> 0110$\,,\, $10001 \> 01010$\,,\, $00101 \> 10000$\,,\, $00111 \>
	10010$\,,\, and $01011 \> 10100$ qualify as new rules.	 Theoretically, one
	Their run the procedure on arbitrarily long strings to generate all possible
	comparisons.  As of now, this is the only reliable way to enumerate
	comparisons that are not consequences of each other.

	Executing that, one might find interesting patterns among those comparisons.
	For instance, $10b01$ seems to be $\> 01b10$ for any binary string $b$.
	After some reductions, one sees that $10 0^m 01 \> 01 0^m 10$ seems to be a
	set of rules that are not consequences of each other.  Can we prove this
	for all $m$?

\section{Inverse of Density Evolution}\label{sec:devolve}

	To motivate why we need the inverses of $W \mapsto W^0$ and $W \mapsto W^1$,
	consider the following.  We know from running the deterministic procedure
	that $10 0^m 01 \> 01 0^m 10$ holds for $m = 0, 1, \dotsc, 10$.  But we want
	it to hold for all positive integers $m$.  Wouldn't it be convenient if
	there is a comparison $c \> d$ that, when combined with $10 0^m 01
	\> 01 0^m 10$, gives
	\begin{align}
		10 0^{m+1} 01 & = c 10 0^m 01 \label{equ:c}\\
		& \> d 01 0^m 10 = 01 0^{m+1} 10? \label{equ:d}
	\end{align}
	What should $c$ and $d$ be to make this happen?

	By reverse-engineering \eqref{equ:c} and \eqref{equ:d}, we see that $c$
	``equals'' $100 0^{-1} 1^{-1}$ and $d$ ``equals'' $010 1^{-1} 0^{-1}$.  The
	inverses $0^{-1}$ and $1^{-1}$ can be made formal by appointing $I_2(x)
	\coloneqq \sqrt x$ the inverse function of $I_0$ and $I_3(x) \coloneqq 1 -
	\sqrt{1 - x}$ the inverse function of $I_1$.  We use $2$ and $3$ in place of
	$0^{-1}$ and $1^{-1}$ to ease the notation.  These are \emph{density
	devolution}; if $W$ is a BEC with capacity $x$, then $U^0 = V^1 = W$, where
	$U$ and $V$ are BECs with capacities $I_2(x)$ and $I_3(x)$, respectively.

	Define $I_{q_1 q_2 \dotsm q_n} (x) \coloneqq I_{q_2 \dotsm q_n} (I_{q_1}
	(x))$ for any quaternary string $q_1 q_2 \dotsm q_n \in \{0, 1, 2, 3\}$.
	Note that $02 = 20 = 13 = 31 = \epsilon$.
 
	\begin{definition}
		For any quaternary strings $p$ and $q$, we say $p \> q$ if $I_p(x) \geq
		I_q(x)$ for all $x \in [0, 1]$.
	\end{definition}

	We find generalizations of the old lemmas to the new alphabet as well as new
	lemmas that only make sense with the new alphabet.  Some proofs are omitted.

	\begin{lemma}
		$I_0$, $I_1$, $I_2$, and $I_3$ are strictly monotonically increasing and
		map $[0, 1]$ bijectively onto $[0, 1]$.  In fact, all $I_q$ are.
	\end{lemma}

	\begin{lemma}
		$\>$ still forms a preorder on quaternary strings.  $p \> q$ and $r \>
		s$ still imply $pr \> qs$ and $\tilde q \> \tilde p$.  Here, the
		complement (denoted by tilde) of $2$ and $3$ are each other.
	\end{lemma}

	\begin{lemma}
		$p \> q$
		iff $p q^{-1} \> \epsilon$
		iff $q^{-1} p \> \epsilon$
		iff $q^{-1} \> p^{-1}$
		iff $\epsilon \> p^{-1} q$
		iff $\epsilon \> q p^{-1}$.
		Here, $\epsilon$ represents the empty string.
	\end{lemma}

	\begin{IEEEproof}
		Let $r \coloneqq s \coloneqq q^{-1}$ to prove the first ``only if''.
		The rest are similar.
	\end{IEEEproof}

	\begin{lemma}
		$q^2 \> \epsilon$ iff $q \> \epsilon$.
	\end{lemma}

	\begin{IEEEproof}
		$q^2 \> \epsilon$ implies $q \> q^{-1}$.  Note that the plot of $I_q$
		and the plot of $I_{q^{-1}}$ are symmetric with respect to the diagonal
		that passes $(0, 0)$ and $(1, 1)$.  As $q \> q^{-1}$ implies that the
		former plot lies entirely above the later plot, the former plot lies
		entirely above the diagonal.  This is saying $q \> \epsilon$.
	\end{IEEEproof}

	So far, we have explored some basic properties concerning the quaternary
	alphabet.  But let's not forget that the goal is to make \eqref{equ:c} and
	\eqref{equ:d} (and presumably more of that form) mathematically sound.  How
	exactly can we prove $c \coloneqq 10023 \> 01032 \eqqcolon d$?

\section{Field Extensions of Univariate Polynomials}\label{sec:field}

	We aim to generalize the procedure in Section~\ref{sec:2ary} to one that can
	decides whether $p \> q$ for all quaternary strings $p$ and $q$.  This
	sections prepares some building blocks for such a procedure.

	Let $\QQ$ be the set of rational numbers.
	Let $\QQ[x]$ be the set of univariate polynomials in variable $x$
	whose coefficients are in $\QQ$.
	Let $\KK \coloneqq \Frac(\QQ[x])$ be the fraction field of $\QQ[x]$.
	Let $\CC$ be the set of complex numbers.
	Let $\CC^\CC$ be the set of functions from $\CC$ to $\CC \cup \{\infty\}$
	that take infinity finitely many times.

	A complex number $\xi \in \CC$ is called an \emph{algebraic number} if there
	exists a polynomial $f \in \QQ[x]$ with rational number coefficients such
	that $f(\xi) = 0$.  Since our quaternary alphabet involves extracting the
	square root of a polynomial, we want to extend the concept of algebraic
	numbers to polynomials.

	\begin{definition}
		We call any function $\varphi \in \CC^\CC$ an \emph{algebraic function}
		if there exists a bivariate polynomial $G \in \QQ[x, y]$ with rational
		number coefficients such that $G(\xi, \varphi(\xi)) = 0$ for all $\xi
		\in \CC$ except for the case $\varphi(\xi) = \infty$.  We denote the set
		of algebraic functions by $\AAFF$.
	\end{definition}

	Clearly, $\QQ \subset \QQ[x] \subset \KK \subset \AAFF \subset \CC^\CC$.


	\begin{lemma}
		Algebraic functions are closed under addition and multiplication.  That
		is, $\varphi + \psi, \varphi\psi \in \AAFF$ given $\varphi, \psi \in
		\AAFF$.
	\end{lemma}

	\begin{IEEEproof}
		We use a routine argument from the field theory.  Readers are encouraged
		to skip should they find this boring.

		Let $\KK[\varphi, \psi]$, as a subset of $\CC^\CC$, be the collection of
		functions of the form \[ \sum_{i,j} \frac{p_{i,j}}{q_{i,j}} \varphi^i
		\psi^j, \] where the sum is finite and $p_{i,j}, q_{i,j} \in \QQ[x]$.
		$\KK[\varphi, \psi]$ forms a vector space over $\KK$ because it is
		closed under addition and scalar-multiplication by an element of $\KK$.

		Suppose the degree of $\varphi$ over $\KK$ is $d$ and the degree of
		$\psi$ over $\KK$ is $e$.  That is to say, there are relations
		\[
			\sum_{i=0}^d g_i \varphi^i = 0, \qquad
			\sum_{i=0}^e h_j \psi^j = 0, \qquad
		\]
		where $g_i, h_j \in \QQ[x]$.  Then we can express $\varphi^d$ as a
		combination of $\varphi^i$ for $i < d$.  Likewise, we can express $\psi^e$
		as a combination of lower power terms.  Hence $\KK[\varphi, \psi]$ is a
		finite dimensional vector space over $\KK$.

		This implies that $\KK[\varphi\psi]$, as a vector subspace of
		$\KK[\varphi, \psi]$, is of finite dimension over $\KK$.  Hence $1,
		\varphi\psi, (\varphi\psi)^2, \dotsc, (\varphi\psi)^{d+e}$ are not
		linearly independent over $\KK$.  There must exist a linear equation
		relating the powers of $\varphi\psi$, which witnesses the fact that
		$\varphi\psi$ is an algebraic function.

		For $\varphi + \psi$, the same argument applies.
	\end{IEEEproof}

	\begin{lemma}
		Algebraic functions are closed under extracting the $m$th roots.  That
		is, $\varphi \in \AAFF$ if the function $\varphi \in \CC^\CC$ is such
		that $\varphi^m \in \AAFF$ for some positive integer $m$.
	\end{lemma}

	\begin{IEEEproof}
		If $G(x, \varphi^m(x)) = 0$, then $H(x, y) \coloneqq G(x, y^m)$ is a
		bivariate polynomial such that $H(x, \varphi(x)) = 0$.
	\end{IEEEproof}

	\begin{definition}
		A \emph{root} of an algebraic function $\varphi \in \AAFF$ is a complex
		number $\xi \in \CC$ such that $\varphi(\xi) = 0$.
	\end{definition}

	\begin{lemma}
		A root of an algebraic function is an algebraic number.
	\end{lemma}

	\begin{IEEEproof}
		If $\xi \in \CC$ is such that $\varphi(\xi) = 0$, then $f(x) \coloneqq
		G(x, 0)$ is a polynomial in $x$ such that $f(\xi) = G(\xi, 0) = G(\xi,
		\varphi(\xi)) = 0$.  Note that $f(x)$ is a polynomial with
		rational-number coefficients so $\xi$ is an algebraic number.
	\end{IEEEproof}

	\begin{theorem}
		Suppose $\xi \in [0, 1]$ and $p$ and $q$ are quaternary strings.  If
		$I_p(\xi) = I_q(\xi)$, then $\xi$ is an algebraic number.
	\end{theorem}

	\begin{IEEEproof}
		The building blocks of $I_q$ are squaring, subtracting from $1$, and
		extracting the square root.  We have seen that all three operations map
		an algebraic function to an algebraic function.  Hence $I_p(x) - I_q(x)$
		is an algebraic function.  Hence $\xi$ is an algebraic number.
	\end{IEEEproof}

\section{Algebraic Function Inequality Problem}\label{sec:4ary}

	Thanks to the previous section, we can verify any inequality of the form
	\begin{equation}
		I_p(x) \geq I_q(x) \qquad \text{for all } x \in [0, 1],
		\label{ine:4ary}
	\end{equation}
	where $p$ and $q$ are quaternary strings.
	Here is how.

\subsection{The ``in principle, we can'' part}

	Let $\varphi(x) \coloneqq I_p(x) - I_q(x) \in \CC^\CC$.  To safely extend the
	domain of $I_p - I_q$ to $\CC$, we always select the square root in $[0, 1]$
	if there is any, and select an arbitrary one otherwise.

	By the proof of that any root of $\varphi$ is an algebraic number, there is
	an algorithm that outputs a polynomial $f \in \QQ[x]$ that evaluates any root
	of $\varphi$ to zero.  That is to say, to check if $[0, 1]$ contains any
	root of $\varphi$, it suffices to check if any root of $f$ in $[0, 1]$
	happens to be a root of $\varphi$.

	Now, let the roots of $f$ that are also in $[0, 1]$ be, in ascending order,
	$0 = \xi_0 < \xi_1 < \xi_2 < \dotsb < \xi_m = 1$.  It suffices to check if
	$\varphi((\xi_j + \xi_{j+1}) / 2)$ are all positive.  It might seem that
	this step will cost a lot because we need precise roots of $f$.  Luckily, a
	root-isolating algorithm will give us disjoint intervals that each contains
	an isolated root.  Thus, instead of $(\xi_j + \xi_{j+1}) / 2$, checking the
	endpoints of the intervals is equally valid.

\subsection{The ``thanks to Galois theory, we'd better'' part}

	But how exactly are we going to find the polynomial $f$ given quaternary
	strings $p$ and $q$?  Instead of relying on the lemma that $\varphi\psi$ and
	$\varphi + \psi$ are finite-dimensional over $\KK$ but the linear equations
	are implicit.  The following is what we actually do to work out $f$ more
	efficiently.

	First and foremost, instead of $I_p(x) - I_q(x)$, we will be working on
	$\varphi(x) \coloneqq I_{pq^{-1}} (x) - x$.  If $\varphi$ contains no square
	root, then it is a polynomial and we have detailed how to handle polynomials
	in Section~\ref{sec:2ary}.  Otherwise, $\varphi$ is of the form
	\[ \varphi(x) = H \bigl( x, \sqrt{\psi(x)} \bigr), \]
	where $H \in \QQ[x, y]$ and $\psi \in \AAFF$.  We see that, if $\varphi(\xi)
	= 0$, then
	\[
		H \bigl( x, \sqrt{\psi(\xi)} \bigr) \cdot
		H \bigl( x, -\sqrt{\psi(\xi)} \bigr) = 0
	\]
	Note that $H(x, \sqrt y) \cdot H(x, -\sqrt y) \in \QQ[x, y]$.  So we can
	eliminate one layer of square root by replacing $\varphi$ with
	\begin{equation}
		H \bigl( x, \sqrt{\psi(x)} \bigr) \cdot
		H \bigl( x, -\sqrt{\psi(x)} \bigr).
		\label{for:conjugate}
	\end{equation}
	Repeating the same procedure, we can eliminate all square roots
	introduced by symbols $2$ and $3$ in $pq^{-1}$.
	And the resulting product of $H$'s will be our $f$.

	As an example, suppose we begin with
	\[ \varphi(x) \coloneqq 1 - x - \sqrt{2 - \sqrt{3 - x^2}}. \]
	This means $H(x, y) = 1 - x - y$ and $\psi(x) = 2 - \sqrt{3 - x^2}$.  So
	\eqref{for:conjugate} becomes
	\[ (1 - x)^2 - (2 - \sqrt{3 - x^2}). \]
	Proceed to the next step, $H(x, y)$ becomes $(1 - x)^2 - 2 + y$ and
	$\psi(x)$ becomes $3 - x^2$.  So \eqref{for:conjugate} becomes
	\begin{equation}
		((1 - x)^2 - 2)^2 - (3 - x^2).
		\label{for:f}
	\end{equation}
	Now this is the polynomial $f$ we want.

	Remark: the idea presented in this subsection can be summarized as
	``multiplying all Galois-conjugates together.'' To elaborate, since
	$\varphi$ is algebraic over $\KK$, there exists Galois actions $\sigma_1,
	\sigma_2, \dotsc, \sigma_d$ that sends $\varphi$ to other functions that
	satisfies the very same equation satisfied by $\varphi$.  The product of all
	conjugates must be in the base field: $f \coloneqq \sigma_1(\varphi)
	\sigma_2(\varphi) \dotsm \sigma_d(\varphi) \in \KK$.  But one of the factors
	is $\varphi$; so $\varphi(\xi) = 0$ implies $f(\xi) = 0$.  In the example
	above, the conjugates are
	\begin{gather*}
		1 - x - \sqrt{2 - \sqrt{3 - x^2}} \,,\,
		1 - x + \sqrt{2 - \sqrt{3 - x^2}} \,,\\
		1 - x - \sqrt{2 + \sqrt{3 - x^2}} \,,\,
		1 - x + \sqrt{2 + \sqrt{3 - x^2}}.
	\end{gather*}
	Their product is the same $f$ we conclude in \eqref{for:f}.

	\begin{theorem}
		The procedure described in this section decides inequalities of the form
		\eqref{ine:4ary}.
	\end{theorem}

\section{The TBM Partial Order}\label{sec:tbm}

\subsection{Prefix TBM}

	Inspired by the proposed proof of $10 0^m 01 \> 01 0^m 10$, i.e.,
	\eqref{equ:c} and \eqref{equ:d}, we make the following definition.

	\begin{definition}
		Fix quaternary strings $\alpha$ and $\beta$.  We write $\alpha \tbm>
		\beta$ if, for any quaternary strings $y$ and $z$,
		\[
			\alpha y \> \beta z \qquad \text{implies} \qquad
			\alpha 0y \> \beta 0z \quad\text{and}\quad \alpha 1y \> \beta 1z.
		\]
	\end{definition}

	We call $\tbm>$ a TBM relation.  TBM stands for tunnel boring machine, a
	machine that extends existing tunnels, inspired by the potential usage in
	\eqref{equ:c} and \eqref{equ:d}.  As a starter, we can prove $1 \tbm> 0$
	fairly easily: if $1y \> 0z$, then $10y \> 01y \> 00z$ and $11y \> 10z \>
	01z$.

	Another example: Suppose $10023 \> 01032$ holds, then $10y \> 01z$ implies
	$100y = (10023)(10y) \> (01032)(01z) = 010z$.  Take the bitwise complement:
	$10023 \> 01032$ implies $10123 \> 01132$; then $10y \> 01z$ implies $101y =
	(10123)(10y) \> (01132)(01z) = 011z$.  Therefore, $10023 \> 01032$ implies
	$10 \tbm> 01$.

	As it gets more complicated, we see the necessity to develop a machine that
	can automate this away.

	\begin{lemma}
		$\alpha \tbm> \beta$ if and only if both $\alpha 0 \alpha^{-1} \> \beta
		0 \beta^{-1}$ and $\alpha 1 \alpha^{-1} \> \beta 1 \beta^{-1}$ hold.
	\end{lemma}

	\begin{IEEEproof}
		Let's prove the ``if'' part.  Suppose $\alpha 0 \alpha^{-1} \> \beta 0
		\beta^{-1}$ holds and suppose $\alpha y \> \beta z$, then $(\alpha 0
		\alpha^{-1}) (\alpha y) \> (\beta 0 \beta^{-1}) (\beta z)$, implying
		$\alpha 0y \> \beta 0z$.  For $\alpha 1 \alpha^{-1} \> \beta 1
		\beta^{-1}$, the same argument applies.

		Let's prove the ``only if'' part.  Let $(y, z) \coloneqq (\alpha^{-1},
		\beta^{-1})$, then $\alpha y \> \beta z$ definitely holds as both sides
		are empty.  But then that implies that $\alpha 0y \> \beta 0z$, which is
		$\alpha 0 \alpha^{-1} \> \beta 0 \beta^{-1}$.  Likewise, $\alpha 1y \>
		\beta 1z$ is just $\alpha 1 \alpha^{-1} \> \beta 1 \beta^{-1}$.
	\end{IEEEproof}

	\begin{lemma}
		$\tbm>$ is a preorder.
	\end{lemma}

	\begin{IEEEproof}
		Reflexivity: Observe that $\alpha 0 \alpha ^{-1} \> \alpha 0
		\alpha ^{-1}$ and $\alpha 1 \alpha ^{-1} \> \alpha 1 \alpha ^{-1}$.
		Transitivity: Use $\alpha 0 \alpha^{-1} \> \beta 0 \beta^{-1} \> \gamma
		0 \gamma^{-1}$ and $\alpha 1 \alpha^{-1} \> \beta 1 \beta^{-1} \> \gamma
		1 \gamma^{-1}$.
	\end{IEEEproof}

	\begin{lemma}
		$\alpha \tbm> \beta$ iff $\tilde \beta \tbm> \tilde \alpha$ iff
		$\beta^{-1} \alpha \tbm> \epsilon$ iff $\epsilon \tbm> \alpha^{-1}
		\beta$.
	\end{lemma}
	
	\begin{theorem}\label{thm:finer}
		$\alpha \tbm> \beta$ implies $\alpha \> \beta$.  Note: the converse is
		not true as counterexamples are found.
	\end{theorem}
	
	\begin{IEEEproof}
		This is a sketch of proof.  See the appendix for more details.
		
		If $\alpha \> \beta$ is not true, then there exists $\xi \in [0, 1]$
		such that $I_\alpha(\xi) < I_\beta(\xi)$.  Pick a binary string $c$ such
		that $I_{\alpha c} (\xi)$ is very, very close to $0$, so close that
		$I_{\alpha c \alpha^{-1}} (\xi) < 1/2$.  At the same time, make
		$I_{\beta c} (\xi)$ very, very close to $1$, so close that $I_{\beta c
		\beta^{-1}} (\xi) > 1/2$.  Then $I_{\alpha c \alpha^{-1}} (\xi) <
		I_{\beta c \beta^{-1}} (\xi)$, which contradicts $\alpha c \alpha^{-1} \>
		\beta c \beta^{-1}$.  So $\alpha \> \beta$ must be true.
	\end{IEEEproof}

\subsection{Some rules of rules}

	Since checking $\alpha 0 \alpha^{-1} \> \beta 0 \beta^{-1}$
	and $\alpha 1 \alpha^{-1} \> \beta 1 \beta^{-1}$
	is straightforward using the procedure described in Section~\ref{sec:4ary},
	we list some TBM relations as follows.

	\begin{itemize}\small
		\item $1 \tbm> \epsilon \tbm> 0$, where $\epsilon$ is the empty string.
		\item $11 \tbm> 10 \tbm> 01 \tbm> 00$.
		\item $111 \tbm> 110 \tbm> 101 \tbm> 011 \tbm> 100 \tbm> 010 \tbm> 001$.
		\item $1111 \tbm> 1110 \tbm> 1101 \tbm>
			1011 \tbm> 1100 \tbm> 1010 \tbm> 1001$.
		\item $1011 \tbm> 0111 \tbm> 1001 \tbm> 0110 \tbm> 1000 \tbm> 0100$.
		\item $0110 \tbm> 0101 \tbm> 0011 \tbm>
			0100 \tbm> 0010 \tbm> 0001 \tbm> 0000$.
	\end{itemize}

	As commented before, $10 \tbm> 01$ is equivalent to $10023 \> 01032$;
	we now know that they both are true.
	Hence, as claimed before,
	we have $10 0^m 01 \> 01 0^m 10$ for any positive integer $m$.
	
	Another example: $100011 \> 010110$ and  $10 \tbm> 01$ imply $10 0^m 0011 \>
	01 0^m 0110$ for any positive integer $m$.


	Yet another example:  $011010 \> 100011$ and $011 \tbm> 100$, imply $011 0^m
	010 \> 100 0^m 011$ for any positive integer $m$.


\subsection{Suffix TBM}

	One might be motivated to make the following definition:
	$\alpha \tbm>' \beta$ if,
	for any quaternary strings $y$ and $z$,
	\[
		y \alpha \< z \beta \qquad \text{implies} \qquad
		y0 \alpha \< z0 \beta \text{ and } y1 \alpha \< z1 \beta.
	\]
	Nevertheless, this is not a new concept,
	because $\alpha \tbm>' \beta$ is equivalent to
	$\beta^{-1} 0 \beta \> \alpha^{-1} 0 \alpha$,
	which is equivalent to $\beta^{-1} \tbm> \alpha^{-1}$.

	The following are some suffix TBM relations we found.
	\begin{itemize}\small
		\begin{minipage}{0.49\linewidth}
			\item $1 \tbm>' \epsilon \tbm>' 0$.
			\item $11 \tbm>' 10 \tbm>' 01 \tbm>' 00$.
			\item $111 \tbm>' 101 \tbm>' 011$.
			\item $110 \tbm>' 011 \tbm>' 001$.
		\end{minipage}
		\begin{minipage}{0.49\linewidth}
			\item $101 \tbm>' 010$.
			\item $110 \tbm>' 100 \tbm>' 001$.
			\item $100 \tbm>' 010 \tbm>' 000$.
		\end{minipage}
	\end{itemize}
	\smallskip

	An example application for suffix TBM: $1000101 \> 0101010$ and $10 \tbm>'
	01$, implies $10001 0^n 01 \> 01010 0^n 10$ for any positive integer $n$.


	Another example application for suffix TBM: $001101 \> 100010$ and and $101
	\tbm>' 010$, imply $001 0^n 101 \> 100 0^n 010$ for any positive integer
	$n$.

	
	To the best of our knowledge, all example inequalities given in this section
	are independent rules.  These are particular examples of family of
	infinitely many rules.  In the next section, we relate families of rules to
	the dominance order of integer partition. 

\section{Integer Partition and Dominance Order}\label{sec:partition}

	For a binary string $a$, let $\pi(a)$ be the integer partition $p_1 + p_2 +
	\dotsb + p_m$ where $p_i$ is the number of zeros to the right of the $i$th
	one in $1a0$.  For instance $\pi(10001)$ is $4 + 4 + 1$.  We claim the
	following theorem.

	\begin{theorem}\label{thm:dominance}
		If $a$ and $b$ are of the same length and weight, and $\pi(a)$ dominates
		$\pi(b)$ as in the theory of integer partition, then $a \> b$.
	\end{theorem}

	\begin{IEEEproof}[Sketch of a proof]
		It suffices to prove that, if $\pi(a)$ covers $\pi(b)$ in the dominance
		order, then $a \> b$.  By a theorem in integer partition, $\pi(a)$
		covers $\pi(b)$ if $a$ and $b$ contain $10 \phi^m 01$ and $01 \phi^m
		10$, respectively, where $\phi \in \{0, 1\}$.  But we already know $10
		\phi^m 01 \> 01 \phi^m 10$: they are consequences of $1001 \> 0110$ and
		$10 \tbm> 01$.
	\end{IEEEproof}

	For a binary string $b$, let $\Lindenmayer(b, 0{\to}01, 1{\to}10)$ be the
	resulting string wherein each zero in $b$ is replaced by $01$ and each $1$
	by $10$.  We claim that there are two more embeddings of the dominance
	order.

	\begin{theorem}\label{thm:more}
		Same assumptions as Theorem~\ref{thm:dominance}.
		Then $\Lindenmayer(a$, $0{\to}01$, $1{\to}10)
		\> \Lindenmayer(b$, $0{\to}01$, $1{\to}10)$.
		What's more, $\Lindenmayer(a$, $0{\to}100$, $1{\to}011)
		\> \Lindenmayer(b$, $0{\to}100$, $1{\to}011)$.
	\end{theorem}

	\begin{IEEEproof}
		It suffices to check $10010110 \> 01101001$ and $1001 \tbm> 0110$ for
		the first inequality, and $011100100011 \> 100011011100$ and $011100
		\tbm> 100011$ for the second inequality.  All of them are checked on a
		personal computer.
	\end{IEEEproof}

\section{Conclusions}

	This paper addresses the problem of determining if one synthetic channel is
	always more preferred over another synthetic channel.  We use a
	root-counting algorithm to demonstrate how to determine if a polynomial is
	nonnegative over $[0, 1]$.  We then extend that to a procedure that
	determines if an algebraic function is nonnegative over $[0, 1]$.  The
	latter becomes a tool that can prove infinitely many rules in one go.

\bibliographystyle{IEEEtran}
\bibliography{Polder-Conjugate-25.bib}

\begin{thebibliography}{10}
\providecommand{\url}[1]{#1}
\csname url@samestyle\endcsname
\providecommand{\newblock}{\relax}
\providecommand{\bibinfo}[2]{#2}
\providecommand{\BIBentrySTDinterwordspacing}{\spaceskip=0pt\relax}
\providecommand{\BIBentryALTinterwordstretchfactor}{4}
\providecommand{\BIBentryALTinterwordspacing}{\spaceskip=\fontdimen2\font plus
\BIBentryALTinterwordstretchfactor\fontdimen3\font minus
  \fontdimen4\font\relax}
\providecommand{\BIBforeignlanguage}[2]{{%
\expandafter\ifx\csname l@#1\endcsname\relax
\typeout{** WARNING: IEEEtran.bst: No hyphenation pattern has been}%
\typeout{** loaded for the language `#1'. Using the pattern for}%
\typeout{** the default language instead.}%
\else
\language=\csname l@#1\endcsname
\fi
#2}}
\providecommand{\BIBdecl}{\relax}
\BIBdecl

\bibitem{Ari09}
E.~Arikan, ``Channel polarization: A method for constructing capacity-achieving
  codes for symmetric binary-input memoryless channels,'' \emph{IEEE
  Transactions on Information Theory}, vol.~55, no.~7, pp. 3051--3073, July
  2009.

\bibitem{Ari08}
------, ``A performance comparison of polar codes and reed-muller codes,''
  \emph{IEEE Communications Letters}, vol.~12, no.~6, pp. 447--449, June 2008.

\bibitem{Ari10}
------, ``A survey of reed-muller codes from polar coding perspective,'' in
  \emph{2010 IEEE Information Theory Workshop on Information Theory (ITW 2010,
  Cairo)}, Jan 2010, pp. 1--5.

\bibitem{MHU14}
M.~Mondelli, S.~H. Hassani, and R.~L. Urbanke, ``From polar to reed-muller
  codes: A technique to improve the finite-length performance,'' \emph{IEEE
  Transactions on Communications}, vol.~62, no.~9, pp. 3084--3091, Sep. 2014.

\bibitem{LST14}
\BIBentryALTinterwordspacing
B.~Li, H.~Shen, and D.~Tse, ``A rm-polar codes,'' 2014. [Online]. Available:
  \url{https://arxiv.org/abs/1407.5483}
\BIBentrySTDinterwordspacing

\bibitem{AbY20}
E.~Abbe and M.~Ye, ``Reed-muller codes polarize,'' \emph{IEEE Transactions on
  Information Theory}, vol.~66, no.~12, pp. 7311--7332, Dec 2020.

\bibitem{KKM17}
S.~Kudekar, S.~Kumar, M.~Mondelli, H.~D. Pfister, E.~{\c S}a{\c s}o{\v g}lu,
  and R.~L. Urbanke, ``Reed--muller codes achieve capacity on erasure
  channels,'' \emph{IEEE Transactions on Information Theory}, vol.~63, no.~7,
  pp. 4298--4316, July 2017.

\bibitem{ReP21}
\BIBentryALTinterwordspacing
G.~Reeves and H.~D. Pfister, ``Reed-muller codes achieve capacity on bms
  channels,'' 2021. [Online]. Available: \url{https://arxiv.org/abs/2110.14631}
\BIBentrySTDinterwordspacing

\bibitem{MoT09}
R.~Mori and T.~Tanaka, ``Performance of polar codes with the construction using
  density evolution,'' \emph{IEEE Communications Letters}, vol.~13, no.~7, pp.
  519--521, July 2009.

\bibitem{Sch16}
C.~Sch{\"u}rch, ``A partial order for the synthesized channels of a polar
  code,'' in \emph{2016 IEEE International Symposium on Information Theory
  (ISIT)}, July 2016, pp. 220--224.

\bibitem{BDO16}
M.~Bardet, V.~Dragoi, A.~Otmani, and J.-P. Tillich, ``Algebraic properties of
  polar codes from a new polynomial formalism,'' in \emph{2016 IEEE
  International Symposium on Information Theory (ISIT)}, July 2016, pp.
  230--234.

\bibitem{MHU19}
M.~Mondelli, S.~H. Hassani, and R.~L. Urbanke, ``Construction of polar codes
  with sublinear complexity,'' \emph{IEEE Transactions on Information Theory},
  vol.~65, no.~5, pp. 2782--2791, May 2019.

\bibitem{WuS19}
W.~Wu and P.~H. Siegel, ``Generalized partial orders for polar code
  bit-channels,'' \emph{IEEE Transactions on Information Theory}, vol.~65,
  no.~11, pp. 7114--7130, Nov 2019.

\bibitem{CLM21}
E.~Camps, H.~H. L{\'o}pez, G.~L. Matthews, and E.~Sarmiento, ``Polar decreasing
  monomial-cartesian codes,'' \emph{IEEE Transactions on Information Theory},
  vol.~67, no.~6, pp. 3664--3674, June 2021.

\bibitem{DrC21}
\BIBentryALTinterwordspacing
V.-F. Dr{\u a}goi and G.~Cristescu, ``Bhattacharyya parameter of monomial codes
  for the binary erasure channel: From pointwise to average reliability,''
  \emph{Sensors}, vol.~21, no.~9, 2021. [Online]. Available:
  \url{https://www.mdpi.com/1424-8220/21/9/2976}
\BIBentrySTDinterwordspacing

\bibitem{Gei18}
\BIBentryALTinterwordspacing
B.~C. Geiger, ``The fractality of polar and reed--muller codes,''
  \emph{Entropy}, vol.~20, no.~1, 2018. [Online]. Available:
  \url{https://www.mdpi.com/1099-4300/20/1/70}
\BIBentrySTDinterwordspacing

\bibitem{Kah17}
\BIBentryALTinterwordspacing
S.~Kahraman, ``Strange attractor for efficient polar code design,'' 2017.
  [Online]. Available: \url{https://arxiv.org/abs/1708.04167}
\BIBentrySTDinterwordspacing

\bibitem{KOK16}
D.~Kim, K.~Oh, D.~Kim, and J.~Ha, ``Information set analysis of polar codes,''
  in \emph{2016 International Conference on Information and Communication
  Technology Convergence (ICTC)}, Oct 2016, pp. 813--815.

\bibitem{SaM16}
\BIBentryALTinterwordspacing
M.~Sagraloff and K.~Mehlhorn, ``Computing real roots of real polynomials,''
  \emph{Journal of Symbolic Computation}, vol.~73, pp. 46--86, 2016. [Online].
  Available:
  \url{https://www.sciencedirect.com/science/article/pii/S0747717115000292}
\BIBentrySTDinterwordspacing

\bibitem{Yun76}
\BIBentryALTinterwordspacing
D.~Y. Yun, ``On square-free decomposition algorithms,'' in \emph{Proceedings of
  the Third ACM Symposium on Symbolic and Algebraic Computation}, ser. SYMSAC
  '76.\hskip 1em plus 0.5em minus 0.4em\relax New York, NY, USA: Association
  for Computing Machinery, 1976, pp. 26--35. [Online]. Available:
  \url{https://doi.org/10.1145/800205.806320}
\BIBentrySTDinterwordspacing

\end{thebibliography}

\newpage

\appendices

\section{Proof of Theorem~\ref{thm:finer}}

	For any binary string $c_1 c_2 \dotsm c_n$,
	\begin{align*}
		\kern1em&\kern-1em
		(\alpha c_1 \alpha^{-1}) (\alpha c_2 \alpha^{-1})
		\dotsm (\alpha c_n \alpha^{-1}) \\
		& \>
		(\beta c_1 \beta^{-1}) (\beta c_2 \beta^{-1})
		\dotsm (\beta c_n \beta^{-1}).
	\end{align*}
	This holds because each $\alpha c_i \alpha^{-1} \> \beta c_i \beta^{-1}$ is
	a consequence of $\alpha \tbm> \beta$.  Therefore, we will have $\alpha c
	\alpha^{-1} \> \beta c \beta^{-1}$ for all binary strings $c$.

	Next, suppose $\alpha \> \beta$ is not true, i.e., there exists $\xi \in [0,
	1]$ such that $I_\alpha(\xi) < I_\beta(\xi)$.  Let $A \coloneqq
	I_\alpha(\xi)$ and let $B \coloneqq I_\beta(\xi)$.
	
	We now attempt to construct capacity-achieving polar codes over
	BECs with capacities $1 - A$ and $B$, respectively.
	Since $1 - A + B > 1$, i.e., the sum of the capacities
	of these two channels exceed one, the bitwise complement of the information set
	of the former overlaps the information set of the latter provided that the
	block length is large enough.  That is, for sufficiently large block
	lengths, there is some binary string $c$ such that $\tilde c$ is in the
	information set of a polar code over BEC with capacity $1 - A$
	and $c$ is in the information
	set of a polar code over BEC with capacity $B$.

	Recall that polar code has exponentially small error probabilities.
	Therefore, for arbitrarily small $\varepsilon > 0$, we will see
	\[
		I_{\tilde c} (1 - A) > 1 - \varepsilon
		\quad\text{and}\quad
		I_c (B) > 1 - \varepsilon
	\]
	as long as $c$ is long enough and taken from the intersection.  Using $I_0(1
	- x) = 1 - I_1(x)$, the inequalities above are equivalent to 
	\[
		I_{\alpha c} (\xi) = I_c (A) < \varepsilon
		\quad\text{and}\quad
		I_{\beta c} (\xi) = I_c (B) > 1 - \varepsilon.
	\]
	That is, $c$ is such that $I_{\alpha c} (\xi)$ is very close to $0$ and
	$I_{\beta c} (\xi)$ is very close to $1$.  If we choose $\varepsilon$
	to be really, really small, we will see
	\[
		I_{\alpha c \alpha^{-1}} (\xi) < I_{\alpha^{-1}} (\varepsilon) < 1/2
	\]
	and
	\[
		I_{\beta c \beta^{-1}} (\xi) > I_{\beta^{-1}} (1 - \varepsilon) > 1/2.
	\]
	This contradicts the fact that $\alpha c \alpha^{-1} \> \beta c \beta^{-1}$,
	so it must be the case that $\alpha \> \beta$ is true.  This finishes the
	proof.

	Consider the preceding proof similar to \cite[Proposition~1]{WuS19}.

\end{document}